\documentclass{elsart}
\usepackage{epsfig}
\usepackage{amssymb}
\begin{document}
\begin{frontmatter}
\title{Studying nonlinear effects on the early stage of phase ordering
using a decomposition method}
\author[label1]{M.I.M. Copetti}
\author[label2]{$\!$, G. Krein}
\author[label3]{$\!$, J.M. Machado}
\author[label2]{and R.S. Marques de Carvalho}
\address[label1]{LANA, Departamento de Matem\'atica, Universidade
Federal de Santa Maria \\
97119-900 Santa Maria, RS, Brazil}
\address[label2]{Instituto de F\'{\i}sica Te\'orica, Universidade
Estadual Paulista \\
Rua Pamplona, 145, 01405-900 S\~ao Paulo, SP - Brazil}
\address[label3]{Departamento de Ci\^encia da Computa\c{c}\~ao e
Estat\'{\i}stica\\
Instituto de Bioci\^encias, Letras e Ci\^encias Exatas,
Universidade Estadual Paulista \\
Rua Crist\'ov\~ao Colombo, 2265, 15054 S\~ao Jos´\'e do Rio Preto
- SP - Brazil}

\begin{abstract}
Nonlinear effects on the early stage of phase ordering are studied
using Adomian's decomposition method for the Ginzburg-Landau
equation for a nonconserved order parameter. While the long-time
regime and the linear behavior at short times of the theory are
well understood, the onset of nonlinearities at short times and
the breaking of the linear theory at different length scales are
less understood. In~the Adomian´s decomposition method, the
solution is systematically calculated in the form of a polynomial
expansion for the order parameter, with a time dependence given as
a series expansion. The method is very accurate for short times,
which allows to incorporate the short-time dynamics of the
nonlinear terms in a analytical and controllable~way.
\end{abstract}


\end{frontmatter}

\section{Introduction}
\label{intro} The study of phase ordering kinetics is of
fundamental importance to the understanding of nonequilibrium
dynamics of phase transitions and is of interest in diverse
branches of physics, ranging from cosmology and elementary
particle physics to the different areas of condensed matter
physics. The typical situation for phase ordering occurs when a
system consisting of a two-phase mixture in a
homogeneous phase is rapidly driven across the critical
coexistence temperature $T_c$ into a nonequilibrium state.
Fluctuations around the initial homogeneous state will develop and
the system will ultimately break into domains of different phases
in space, forming a new equilibrium state.

The dynamics of the time evolution towards equilibrium is a
much-studied problem is nonequilibrium statistical mechanics - for
reviews see Refs.~\cite{rev1} and \cite{rev2}. If the order
parameter $\phi$ that characterizes the phases of the system is
not conserved, the time evolution of $\phi$ is usually described
by the phenomenological Ginzburg-Landau (GL) equation~\cite{rev2}
\begin{equation}
\frac{\partial\phi(\vec{x},t)}{\partial t } = - M \; \frac{\delta
F[\phi(\vec{x},t)]}{\delta \phi} + \zeta(\vec x, t)\, ,
\label{gle}
\end{equation}
where $F[\phi]$ is the coarse-grained free energy functional, and
$M$ is a constant mobility, and $\zeta(\vec x, t)$ is a noise term
that mimics the thermal fluctuations. For the purposes of the
present paper, we take for $F[\phi]$ the usual Landau-Ginzburg
double-well form
\begin{equation}
F[\phi] = \int d^3 x \, \left[ R^2\,
\frac{1}{2}\left(\vec{\nabla}\phi\right)^2 + \frac{1}{2} \,
\epsilon \,\phi^2 +  \frac{1}{4}\, u\, \phi^4 \right]. \label{F}
\end{equation}
Here, $R^2$ is a measure of the interaction range, $\epsilon =
(T-T_c)/T_c$ is the reduced temperature, and $u$ is a positive
constant. In a situation of a temperature quench, $\epsilon =
-|\,\epsilon|$, the equilibrium (long time) values of the order
parameter correspond to $\phi^2 = |\,\epsilon|/u$, the binodal
points. The unstable (spinodal) region corresponds to $\phi^2 \leq
|\,\epsilon|/(3u)$. There are several time regimes associated with
the time evolution towards equilibrium. When $\phi \approx 0$ at
$t=0$, one has the linear regime. The nonlinear terms in
Eq.~(\ref{gle}) are small for small $t$ and can be neglected; the
equation becomes linear~\cite{CH}. In this case, the equation can
be integrated very easily and the (noiseless) solution can be
written as
\begin{equation}
\phi(\vec{x},t) = \int d^3 k \; \exp \left[i\vec{k}\cdot\vec{x} -
M(R^2 k^2 + \epsilon ) \, t \right]\,
\widetilde{\phi}(\vec{k}),\hspace{0.5cm} \rm{for}\;\;\;t\approx 0,
\label{sol-lin}
\end{equation}
where $\widetilde{\phi}(\vec{k})$ is the Fourier transform of
$\phi(\vec{x},t=0)$. In a situation of a temperature quench, there
is an exponential growth of the order parameter for wave-lengths
corresponding to $k^2 \lesssim |\,\epsilon|/R^2$. Because of the
fast exponential growth for long wave-lengths the nonlinearities
rapidly set in, the growth process is slowed down and for longer
times interfaces start to be produced, and the linear theory
cannot be employed anymore. The time scale that characterizes the
onset of the nonlinear regime depends on $M$ and $R^2$. For much
longer time scales, one will reach the final stage of the
coarsening process and the reach of equilibrium.

In the present paper we are mostly interested in the study of the
breakdown of the linear approximation at early stage of the phase
ordering. In particular, we are interested in studying systems
with initial configurations that consist of domains where the
order parameter is not necessarily small. This situation occurs
for example if the initial state of the system consists of domains
composed of metastable matter, i.e. $|\,\epsilon|/u \ge \phi^2 \ge
|\,\epsilon|/(3u)$. Our aim here is to provide a semi-analytical
method to study the short-time evolution of nonlinearities in the
Ginzburg-Landau equation for such situations. The motivation is of
course to obtain an understanding of the time and length scales of
the breakdown of the linear approximation in a systematic and
controllable way, without resorting to some sort of discretization
methods which can easily result in massive numerical computation.

Specifically, we propose to employ Adomian's decomposition
method~\cite{adom} to represent the solution of Eq.~(\ref{gle}) in
terms of a functional series expansion. The method has been used
in several instances~\cite{wazwaz} and more recently it has been
used for the KdV equation~\cite{KdV} and the coupled system of the
Schr\"odinger-KdV equations~\cite{KdV-Sch}. For the GL equation
the use of the method has been outlined in Ref.~\cite{adom-gl},
but no explicit solutions were obtained nor comparisons with
complete solutions were done. In the present letter we show that
the decomposition method can be a useful analytical tool to
investigate the dynamics of the early stage of phase ordering
kinetics. We focus primarily on situations were the linear theory
certainly fails, as in the case when the initial state consists of
large domains with metastable matter. Since the early works of
Ref.~\cite{early}, the study of the early stage of phase ordering
has gained renewed interest in recent
years~\cite{short1}-\cite{short6}. Another direction that the
decomposition method can be useful is in the study of the scaling
behavior at early stages of phase ordering~\cite{scal-early}, a
subject of current interest for thin films roughness
studies~\cite{scal-curr}.

In this first application, to illustrate the applicability and
reliability of the method for the GL equation, we use a simple
double-well free energy functional and present numerical results
for the one-dimensional case only. The initial condition is chosen
to be an oscillatory function to mimic a configuration of domains
with large order parameter. A more detailed investigation at
higher dimensions and different free-energies and initial
configurations is reserved for a future publication. In the next
Section we explain the decomposition method for the GL equation
and obtain the lowest-order terms of the series expansion. In
Section~\ref{results} we show that in the linear approximation the
series can be explicitly summed and gives the exact result. For
the nonlinear regime we use lowest-order terms of decomposition to
compare results with the numerical solution of the GL equation.
Our conclusions and future perspectives are presented in
Section~\ref{concl}.

\section{The decomposition method for the Ginzburg-Landau equation}
\label{decomp_met}
For the free energy of the form given in Eq.~(\ref{F}), the
Ginzburg-Landau equation (we neglect the noise term since in a
quench situation $\epsilon < 0$ the thermal fluctuations being
proportional to $\sqrt{T}$ become smaller after the quench) can be
written as
\begin{equation}
\frac{\partial\phi(\vec{x},t)}{\partial t } = \left( \gamma \,
\nabla^2 + \beta \right) \phi(\vec{x},t) - \alpha \,
A[\phi(\vec{x},t)], \label{che-expl}
\end{equation}
where $A[\phi]$ is the nonlinear term
\begin{equation}
A[\phi(\vec{x},t)] = \phi^3(\vec{x},t)\,, \label{def-A}
\end{equation}
and $\gamma = MR^2, \beta = M |\,\epsilon|$ and $\alpha = Mu$.
Adomian's decomposition method consists in expressing $\phi$ in
the form of a functional series expansion, where the terms of the
expansion are determined recursively. Specifically, the iterative
procedure consists in writing $\phi(\vec x,t)$ and the nonlinear
term $A[\phi]$ as the expansions
\begin{equation}
\phi(x,t)=\sum_{n=0}^{\infty}\phi_n(x,t)\,,\hspace{1.0cm}
A[\phi]=\sum_{n=0}^{\infty}A_n[\phi_0,\phi_1,\cdots,\phi_n]\,,
 \label{expansions}
\end{equation}
where the $A_n$'s are the Adomian polynomials, which can be
determined from
\begin{equation}
A_n[\phi_0,\phi_1,\dots,\phi_n] =
\frac{1}{n!}\left\{\frac{d^n}{d\lambda^n} A[\phi_0+
\lambda^1\phi_1+\cdots+\lambda^n\phi_n]\right\}_{\lambda=0}\;.
\label{Ans}
\end{equation}
Rewriting Eq.~(\ref{che-expl}) as
\begin{equation}
\phi(x,t) = \phi(x,0) + \left( \gamma \, \nabla^2 + \beta
\right)\,\int_0^t dt^{\prime} \, \phi(x,t^{\prime}) - \alpha
\int_0^t dt^{\prime} \, A[\phi(x,t^{\prime})]\,, \label{phiop}
\end{equation}
and substituting the expansions of Eq.~(\ref{expansions}), one
obtains
\begin{eqnarray}
\sum_{n=0}^{\infty}\phi_n(x,t) &=& \phi(\vec x, 0) + \left(
\gamma\;\nabla^2 + \beta \right) \int_0^t dt^{\prime} \,
\sum_{n=0}^{\infty}\phi_n(x,t^{\prime}) \nonumber \\
&-& \alpha \int_0^t dt^{\prime} \,
\sum_{n=0}^{\infty}A_n[\phi_0(x,t^{\prime}), \phi_1(x,t^{\prime}),
\cdots, \phi_n(x,t^{\prime})]\,.
\end{eqnarray}

If one identifies $\phi_0(x,t) = \phi(x,0) \equiv \phi_0(x)$,
Adomian's iteration procedure consists in matching the $(n+1)$-th
order term on the l.h.s. of this equation to the $n$-th order term
on the r.h.s. as
\begin{eqnarray}
\phi_{n+1}(x,t) &=&  \left( \gamma\;\nabla^2 + \beta
\right)\int_0^t dt^{\prime} \, \phi_n(x,t^{\prime}) \nonumber\\
&-& \alpha \int_0^t dt^{\prime} \, A_n[\phi_0(x,t^{\prime}),
\phi_1(x,t^{\prime}), \cdots, \phi_n(x,t^{\prime}) ]\,.
\label{iter}
\end{eqnarray}

For the case when $A[\phi]$ is a cubic form, the $A_n$'s can be
written in a compact way as
\begin{eqnarray}
A_n[\phi_0,\phi_1,\cdots\phi_n] &=&
\frac{1}{n!}\left\{\frac{d^n}{d\lambda^n}
\sum_{i,j,k=0}^{n}\lambda^{i+j+k}\phi_i\phi_j\phi_k\right\}_{\lambda=0}
\nonumber\\
&=& \sum_{i,j,k = 0}^{n}\theta(i+j+k-n) \,\phi_i\phi_j\phi_k \,,
\end{eqnarray}
where $\theta$ is the Heaviside step function.  The general
expression for the $(n+1)$-th term of the expansion of $\phi(\vec
x, t)$ is given as
\begin{eqnarray}
\phi_{n+1}(\vec{x},t) &=& \int_0^t dt^{\prime} \,
\Bigl[(\gamma\nabla^2 + \beta)\phi_n(\vec{x},t^\prime)
\nonumber\\
&-& \alpha \sum_{i,j,k=0}^{n}\theta(i+j+k-n)\,
\phi_i(\vec{x},t^\prime) \phi_j(\vec{x},t^\prime)
\phi_k(\vec{x},t^\prime) \Bigr]\; . \label{phi_n+1}
\end{eqnarray}

In general, it is not possible to express the solution in closed
form. As will be shown in the next section, when the nonlinear
term is neglected, the series can be summed and the solution is
precisely Eq.~(\ref{sol-lin}). The solution will be given in terms of
a power series expansion in $t$ as
\begin{equation}
\phi_n(\vec x, t) =  \bar \phi_n(\vec x)\;\frac{t^n}{n!}\,,
\label{bphi_n}
\end{equation}
where the $\bar\phi_n$'s are easily obtained from
Eq.~(\ref{phi_n+1}). The first few terms are given by
\begin{eqnarray}
\bar\phi_0(\vec x) &=& \phi_0(\vec x)\,,
\label{barphi1}\\
\bar\phi_1(\vec x) &=&  \left( \gamma\;\nabla^2 + \beta
\right)\, \phi_0(x) - \alpha \, \phi^3_0(\vec x)\,, \\
\bar\phi_2(\vec x) &=&  \left( \gamma\;\nabla^2 + \beta
\right)\,\bar\phi_1(x) - 3\,\alpha\left[\phi^2_0(\vec x)
\, \bar\phi_1 (\vec x)\right]\,, \\
\bar\phi_3(\vec x) &=&  \left( \gamma\;\nabla + \beta
\right)\,\bar\phi_2(x) - 3\,\alpha \left[\phi_0(\vec x) \,
\bar\phi^2_1(\vec x) + \bar\phi_2(\vec x) \,
\phi^2_0(\vec x)\right]\,,\\
\bar\phi_4(\vec x) &=&  \left( \gamma\;\nabla^2 + \beta
\right)\,\bar\phi_3(x) \nonumber \\
&-& \alpha\left[\bar\phi^3_1(\vec x) + 3 \phi^2_0(\vec
x)\bar\phi_3(\vec x) + 6 \phi_0(\vec x) \bar\phi_1(\vec
x)\bar\phi_2(\vec x)\right]\,,\\
\bar\phi_5(\vec x) &=&  \left( \gamma\;\nabla^2 + \beta
\right)\,\bar\phi_4(x) \nonumber \\
&-& \alpha\left[3\bar\phi^2_1(\vec x)\bar\phi_2(\vec x) + 3
\phi^2_0(\vec x)\bar\phi_4(\vec x) + 6 \phi_0(\vec x)
\bar\phi_1(\vec x)\bar\phi_3(\vec x)\right]\,,\\
\bar\phi_6(\vec x) &=&  \left( \gamma\;\nabla^2 + \beta
\right)\,\bar\phi_5(x) \nonumber \\
&-& \alpha\left[3\bar\phi^2_1(\vec x)\bar\phi_3(\vec x) +
3\bar\phi^2_2\bar\phi_1 + 3 \phi^2_0(\vec x)\bar\phi_5(\vec x) + 6
\phi_0(\vec x)\bar\phi_2(\vec x)\bar\phi_3(\vec x) \right]\,.
\label{barphi6}
\end{eqnarray}

In the next section we present explicit solutions for the
one-dimensional case. We obtain the exact result in the linear
approximation and for the full nonlinear equation we compare
results obtained with the decomposition method with the full
numerical solution of the Ginzburg-Landau.

\section{Explicit solutions}
\label{results}

We start considering the linear equation ($\alpha =0$). The $n$-th
term is given by
\begin{equation}
\phi_n(\vec x, t) =  \frac{1}{n!}\left[(\gamma\nabla^2 +
\beta)\,t\right]^n\; \phi_0(\vec x)\,. \label{nth}
\end{equation}
The series can be explicitly summed, giving
\begin{equation}
\phi(\vec x,t) = \sum_{n=0}^{\infty} \phi_n(\vec x,t) =
\exp\left[(\gamma\nabla^2 + \beta)\,t\right]\phi_0(\vec x).
\label{expl-sol}
\end{equation}
Using the Fourier transform of the initial condition and applying
the exponential operator to it, one obtains precisely
Eq.~(\ref{sol-lin}) with $\epsilon = -|\,\epsilon|$.

Next, we consider the nonlinear terms. We use an initial
configuration that mimics a system with domains of phases
characterized by an order parameter that is not small. We also
consider a quench to zero temperature ($\epsilon = - 1 $). For
$u^2 = 1$, we have that the unstable (spinodal) region corresponds
to $-\sqrt{1/3} \leq \phi \leq + \sqrt{1/3}$, and the equilibrium
values for the order parameter are $\phi_{equil} = \pm 1$.
Specifically, we use as an initial configuration a periodic
function of the form
\begin{equation}\label{init_cond}
\phi(x,0) = \phi_0 \, \cos(a\,x)\,.
\end{equation}
In the regions of space in the neighborhood of $x = \pm n\pi/a$,
$n=0,1,\cdots$ the order parameter is close to $\phi_0$; as one
departs from these regions the order parameter starts diminishing
until it reaches zero, independently of the value of $\phi_0$. Note
that the average value of $\phi(x,t)$ is zero over a large region
of space.

It is instructive to take a look at the first few terms of the
$\bar\phi$'s of Eqs.~(\ref{barphi1})-(\ref{barphi6}). We present
the explicit forms of the $\bar\phi_n$ up to $n=4$,
\begin{eqnarray}
\bar\phi_0(x) &=& \phi_0 \, \cos (a\,x), \\
\bar\phi_1(x) &=& \phi_0 \,\cos (a\,x)\, \left[ \left(\beta -
a^2\,\gamma\right) - \alpha\, \phi^2_0\, \cos^2 (a\,x) \right], \\
\bar\phi_2(x) &=& \phi_0\,\cos (a\,x)\, \Bigl[ {\left( \beta -
a^2\,\gamma \right) }^2 - \alpha\,\left( 3\,\beta - 5\,a^2\,\gamma
\right) \, \phi^2_0\, \cos^2 (a\,x) \nonumber \\
&+& 2\,{\alpha}^2\, \phi^4_ 0 \, \cos^4 (a\,x)
- 6\, \alpha\, a^2\,\gamma \, \phi^2_0\, \sin^2 (a\,x) \Bigr], \\
\bar\phi_3(x) &=& \phi_0\,\cos (a\,x)\, \Bigl[ {\left( \beta -
a^2\,\gamma \right) }^3 + {\alpha}^2\, \left( 17\,\beta -
31\,a^2\,\gamma \right) \, \phi^4_0 \, \cos^4 (a\,x)  \nonumber \\
&-& 9\,{\alpha}^3\, \phi^6_0 \, \cos^6 (a\,x)  + 24\, \alpha\,
a^2\, \gamma\, \left( -\beta + 3\,a^2\,\gamma \right) \,
\phi^2_0 \,\sin^2(a\,x) \nonumber \\
&+& \alpha\, \Bigl( -9\,{\beta}^2 + 26\,a^2\,\beta\,\gamma -
33\,a^4\,{\gamma}^2 + 58\,a^2\,\alpha\, \gamma\, \phi^2_0 \,\sin^2
(a\,x) \Bigr)\nonumber \\
&\times& \phi^2_0 \, \cos^2 (a\,x) \,  \Bigr], \\
\bar\phi_4(x) &=& \phi_0\,\cos (a\,x)\, \Bigl\{ {\left( \beta -
a^2\,\gamma \right) }^4 - 3\,{\alpha}^3\, \left(
31\,\beta - 67\,a^2\,\gamma \right) \, \phi^6_0 \, {\cos (a\,x)}^6
\nonumber\\
&+& 40\,{\alpha}^4\,{\phi_0}^8\,{\cos (a\,x)}^8 +
{\alpha}^2\,{\phi_0}^4\,{\cos (a\,x)}^4\, \Bigl[ 71\,{\beta}^2 -
260\,a^2\,\beta\,\gamma \nonumber \\
&+& 409\,a^4\,{\gamma}^2
- 598 \,a^2\,\alpha\,\gamma \, \phi^2_0 \,\sin^2 (a\,x) ] +
348\, \alpha^2\, a^4\, {\gamma}^2\,  \phi^4_0\, \sin^4 (a\,x)
\nonumber \\
&-& 6 \,\alpha\, a^2\, \gamma\, \left( 13\,{\beta}^2 -
66\,a^2\,\beta\,\gamma + 117\,a^4\,{\gamma}^2 \right) \,
\phi^2_0\, \sin^2 (a\,x) \nonumber \\
&+& \alpha\, \phi^2_0 \, \cos^2 (a\,x) \, \Bigl[ -19\,{\beta}^3 +
83\,a^2\,{\beta}^2\,\gamma - 189\,a^4\,\beta\,{\gamma}^2 +
253\,a^6\,{\gamma}^3 \nonumber \\
&-& 2 \, \alpha\, a^2\,\gamma\, \left( -253\,\beta +
929\,a^2\,\gamma \right) \, \phi^2_0 \, \sin^2 (a\,x) \Bigr]
\Bigr\}.
\end{eqnarray}

The effects of the nonlinearity are encoded by the terms
proportional to $\alpha$ and involve powers of $\phi_0$ higher
than two. It seems clear that when $\phi_0$ is small, the higher
powers of $\phi_0$ will give small contributions. We verified
explicitly that at small times one is still in the linear regime
at short times when $\phi_0$ is small and the decomposition method
gives the correct answer for all values of $x$. When $\phi_0$ is
large, the nonlinearities will modify the linear term - in the
regions close to the coordinate points $x = \pm n\pi/a$,
$n=0,1,\cdots$ - at relatively low order $n$ , even at very short
times, as we shall see shortly. Therefore, despite the average value
of $\phi$ is zero over a large region of space, for local
regions with large values of $\phi$ the local approximation breaks
down. This is agreement with the discussions in Refs.~\cite{short2}
and \cite{short3} on the breakdown of the linear approximation
at different lenght scales.

For longer times, it seems that
one has to go to large orders in the expansion to obtain a good
approximation to the solution. The situation can be improved if
one is able to perform a (partial) summation of the series. In the
literature~\cite{pade} Pad\'e approximation has been used with
success to improve the convergence of the series. We are not going
to follow this path here, since we are interested in the
short-time behavior of the solutions.

In order to obtain a quantitative estimate of the effects of the
nonlinear terms, we compare results with a numerical solution of
the GL equation. The numerical solution is obtained using a
semi-implicit finite-difference scheme for the time evolution and
a Fast Fourier Transform for the spatial dependence~\cite{mimc}.
Specifically, we use a domain of length $L$, with node points
$x_j$, $j=0,\cdots, N$, with spacing given by $h=L/N$. The value
of $\phi(x,t)$ at point $x_j$ at time $t_n$ is denoted by
$\phi^n_j$. The spatial and time derivatives are discretized as
\begin{equation}
\frac{\partial^2 \phi(x,t)}{\partial x^2} \longrightarrow
\frac{\phi^n_{j+1} - 2\phi^n_j + \phi^n_{j-1}}{h^2},\hspace{1.5cm}
\frac{\partial \phi(x,t)}{\partial t} \longrightarrow
\frac{\phi^{n}_j - \phi^{n-1}_j}{\Delta t}.
\end{equation}
Next, we write the Fourier series for $\phi^n_j$ and
$U(\phi^n_j)=\beta \phi^n_j - \alpha (\phi^{n}_j)^3$ as
\begin{equation}
\phi^n_j = \sum^{N-1}_{k=0} \exp \left(i \frac{2\pi}{N} j \,
k\right)\, \tilde \phi^n(k),\hspace{0.75cm} U(\phi^n_j) =
\sum^{N-1}_{k=0} \exp\left(i \frac{2\pi}{N} j \, k\right)\,
\widetilde U^n(k).
\end{equation}
Substituting these into the GL equation, one can write the
following semi-implicit equation for $\tilde\phi$ at momentum $k$
and time $n$
\begin{equation}
\tilde \phi^n(k) = \frac{1}{1 + \gamma \lambda_k \, \Delta t}
\left[\tilde \phi^{n-1}(k) + \Delta t \, \widetilde
U^{n-1}(k)\right], \label{num-sol}
\end{equation}
where $\lambda_k = [2 - 2 \cos(2\pi k/N)]/h^2$. This scheme has
been shown to be very efficient previously~\cite{mimc}, and seems
to be adequate for our purposes here.

In Figure~1 we plot the solutions at short times ($t \le 1/M$)
obtained with the decomposition method, for the linear (dashed
lines) and nonlinear (long-dashed lines) approximations, and with
the numerical method just described (solid lines). We present the
results for $x=0$. We use $a= \pi/4$, $L=8$ and $\phi_0 =0.5$ and
$\phi_0 = 0.75$ for the initial condition in
Eq.~(\ref{init_cond}); the first value corresponds to unstable
matter and the the second to metastable matter in the region close
to $x=0$ and $x=L$. Note that the average value of $\phi$ over our
spatial length $L$ is zero (we use periodic boundary conditions).

\vspace{0.5cm}
\begin{figure}
\centerline{\epsfig{file=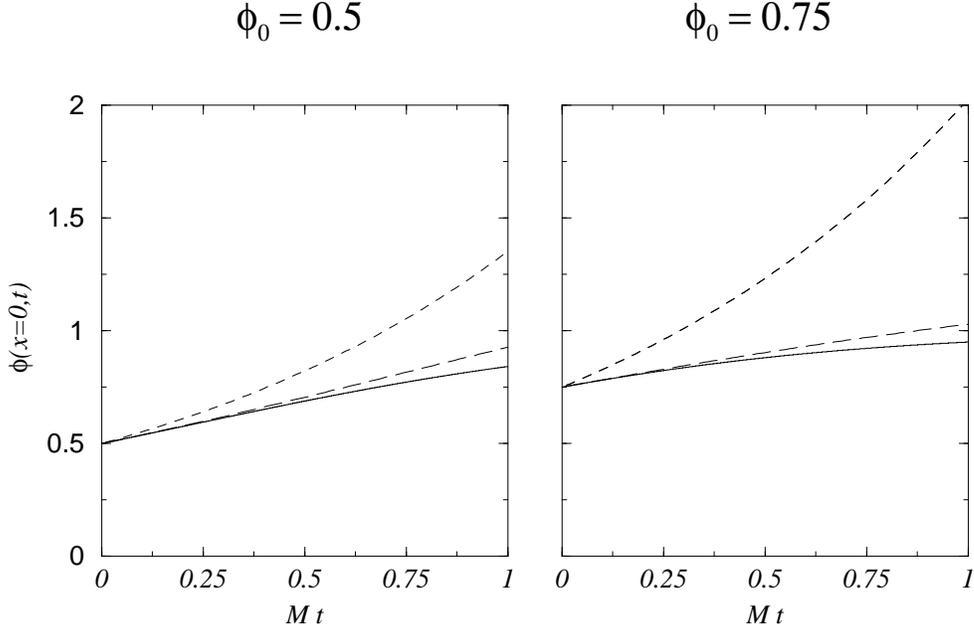,width=13cm} } \caption{The
solutions at short times at $x=0$ of the noiseless GL equation.
The solid line represents the numerical solution using
Eq.~(\ref{num-sol}) and the dashed and long-dashed lines represent
the solutions obtained with the decomposition method up to $6$-th
order, in the linear and nonlinear approximations respectively.}
\label{fig}
\end{figure}

\vspace{0.75cm}
From the plots in the Figure, one clearly sees that the linear
approximation fails dramatically at all time scales. It is also
clear that the decomposition method describes very well the exact
solution at short times, up to the characteristic ``diffusion"
time scale given by $1/M$.

In order to analyze the breakdown of the linear approximation and
the rate of convergence of the expansion we consider the partial
sum
\begin{equation}\label{part_sum}
\phi_N(x,t)= \sum^N_{n=0} \phi_n(x,t).
\end{equation}
The exact solution is obtained for $N\rightarrow \infty$. In
Table~1 we present the ratios of $\phi_N$ to the exact solution
$\phi_{exact}(x=0,t)$ for the linear ($\phi^l_N$) and nonlinear
($\phi^{nl}_N$) approximations for $t = 1/(2M)$ and $t = 1/M$.
One clearly sees again that the decomposition method provides an
excellent approximation to the solution at short times at
relatively low order. For $N=4$, in the nonlinear case, it
approximates the solution to $2.5\%$ accuracy at $t = 1/(2M)$ and
to $10\%$ at $t = 1/M$. The Table also indicates that convergence
becomes slower after such an accuracy has been achieved and a
resummation seems necessary to further improve convergence.

\begin{table} 
\begin{center}
\caption{Ratios of the linear $\phi^l_N(x=0,t)$ and nonlinear
$\phi^{nl}_N(x=0,t)$ approximations to the exact solution
$\phi_{exact}(x=0,t)$, for $t = 1/(2M)$ and $t = 1/M$.}
\vspace{0.25cm}
 \begin{tabular}{c|c|c||c|c}
  \hline
 \multicolumn{5}{c}{\hspace{0.3cm}$t = 1/(2M)$ \hspace{1.25cm} \vline
\hspace{0.9cm} $t = 1/M$} \\
  \hline
  $N$ & $\phi^{l}_N/\phi_{exact}$ & $\phi^{nl}_{N}/\phi_{exact}$
  & $\phi^{l}_N/\phi_{exact}$
  & $\phi^{nl}_{N}/\phi_{exact}$ \\
  \hline\hline
  $1$ & 1.088 & 0.997   & 1.185 & 1.036  \\
  $2$ & 1.178 & 1.031   & 1.478 & 1.146  \\
  $3$ & 1.193 & 1.026   & 1.576 & 1.115  \\
  $4$ & 1.194 & 1.025   & 1.600 & 1.100  \\
  $5$ & 1.195 & 1.025   & 1.605 & 1.100  \\
  $6$ & 1.195 & 1.025   & 1.606 & 1.100  \\
  \hline
 \end{tabular}
\end{center}
\end{table}

\section{Conclusions and Future Perspectives}
\label{concl}

We have proposed to use Adomian's decomposition method to study in
an analytical way the early stage of phase ordering in the context
of the Ginzburg-Landau equation for a nonconserved order
parameter. The method seems to be particularly useful for studying
the time evolution of the order parameter for systems with regions
in space containing metastable matter, where the linear
approximation breaks down. Although we have used a particular free
energy functional and a simple initial configuration, they were
sufficiently representative for the purposes of the present paper.

The results presented here apply to a model A type of equation, with
a double-well free-energy functional. We expect the method to be useful 
for other types of phase ordering models, such as model B, governed 
by a Cahn-Hilliard type of equation. Work in this direction is in progress.
The coupled problem of a conserved and a nonconserved order parameter
(model C) seems to be particularly interesting, since very little
is known about the short time dynamics for nontrivial couplings.

\vspace{1.0cm} \noindent {\bf Acknowledgments}

\noindent Work partially financed by CNPq and FAPESP (Brazilian
agencies).

\end{document}